\documentclass[floatfix,aps,pra,twocolumn,10pt,superscriptaddress]{revtex4-1}
\pdfoutput=1
\usepackage{amsmath}
\usepackage{amsfonts}
\usepackage{amssymb}
\usepackage{graphicx}
\usepackage{tabularx}
\usepackage{verbatim}
\usepackage{color}
\usepackage{algorithm}
\usepackage{todonotes}
\usepackage{siunitx}
\usepackage{algorithm}
\usepackage{algpseudocode}
\usepackage{url}
\usepackage[shortlabels]{enumitem}

\usepackage{mathtools}
\mathtoolsset{showonlyrefs=true}
\newlength\myindent
\begin{document}

\title{Image Acquisition Planning for Earth Observation Satellites with a Quantum Annealer}

\author{Tobias Stollenwerk}
\affiliation {German Aerospace Center (DLR), Linder H\"ohe, 51147 K\"oln, Germany}
\author{Vincent Michaud}
\affiliation {Airbus Defence and Space, 31 rue des Cosmonautes, 31400 Toulouse, France}
\author{Elisabeth Lobe}
\affiliation {German Aerospace Center (DLR), Linder H\"ohe, 51147 K\"oln, Germany}
\author{Mathieu Picard}
\affiliation {Airbus Defence and Space, 31 rue des Cosmonautes, 31400 Toulouse, France}
\author{Achim Basermann}
\affiliation {German Aerospace Center (DLR), Linder H\"ohe, 51147 K\"oln, Germany}
\author{Thierry Botter}
\affiliation {Central Research and Technology, Airbus, 81663 Munich, Germany}

\begin{abstract}
	We present a comparison study of state-of-the-art classical optimisation methods to a D-Wave 2000Q quantum annealer for the planning of Earth observation missions.
The problem is to acquire high value images while obeying the attitude manoeuvring constraint of the satellite.
In order to investigate close to real-world problems, we created benchmark problems by simulating realistic scenarios.
Our results show that a tuned quantum annealing approach can run faster than a classical exact solver for some of the problem instances.
Moreover, we find that the solution quality of the quantum annealer is comparable to the heuristic method used operationally for small problem instances, but degrades rapidly due to the limited precision of the quantum annealer.

\end{abstract}
\maketitle

\section{Introduction}
\label{sec:introduction}
Due to recent hardware developments~\cite{Boixo2018,Versluis2017,Reagor2018,Sete2016}, quantum computing is gaining more and more interest across industry and research domains.
There are indications that heuristic quantum approaches might outperform classical approaches for certain combinatorial optimisation problems~\cite{Denchev2016,Fahri2014}.
The most prominent of these approaches are quantum annealing (QA) and the Quantum Alternating Operator Ansatz (QAOA)~\cite{Hadfield2017}.

In this study, we assess the performance of quantum annealers for the operational planning of an Earth observation satellite. 
Rather than comparing quantum annealing to their classical counterparts~\cite{Denchev2016}, we pursue a more practical and application-driven approach by comparing the performance of quantum annealers to algorithms used in an industry setting. 
There have been multiple studies on solving more or less real-world problems with quantum annealing~\cite{rieffel2015,Venturelli2015,stollenwerkATM2019,stollenwerkFGA2019}.
However, few of them used real world data for benchmarking or compared to state-of-the-art classical solvers which are used operationally in industry.
With this study, we aim for enhancing the insight on the maturity and potential of quantum computing for real-world planning problems; 
in particular from the perspective of potential end-users of quantum computers from industry.

The paper is structured as follows: in Section~\ref{sec:problem_description} we introduce the details of the problem,
before we present a classical solution in Section~\ref{sec:classical_solution};
in Section~\ref{sec:quantum_annealing} we present and analyse the results from the quantum annealer before we compare them to the classical solution in Section~\ref{sec:qa_vs_classical_comparison}.

\section{Problem Description}
\label{sec:problem_description}
In the following, we will describe the Earth observation satellite mission planning problem we are investigating in this work.
For alternative formulations of mission planning problems see \cite{lemaitre2002}.
We are given a set of targets to photograph on the Earth's surface, the \textit{acquisition requests}, together with their commercial value, the \textit{score}.
We focus on a high-resolution, agile satellite in a highly inclined orbit (typically sun synchronous at $\ang{98}$), which provides good coverage of the globe.
The ground swath of a high resolution optical instrument is typically two orders of magnitude lower than the area corresponding to the satellite field of view.
\begin{figure}[b!]
  \centering
  \includegraphics[width=0.85\columnwidth]{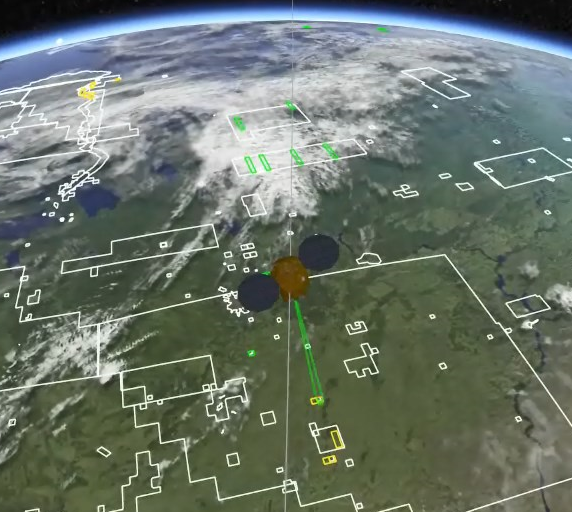}
  \caption{Realistic visualisation of a mission plan where acquisition requests are in white, planned request are in yellow and already acquired request are in green.}
  \label{fig:missionPlanningScale}
\end{figure}
Satellite agility is therefore critical for taking as many pictures as possible.
While the optical instrument remains fixed on the satellite, the whole satellite is manoeuvrable on three axes (roll, pitch and yaw), due to its attitude and orbit control system.
The satellite can therefore manoeuvre during image acquisitions and during transitions between images \cite{lemaitre2002}.
Figure \ref{fig:missionPlanningScale} depicts a realistic scenario visualisation.
The agility of the satellite allows it to start the acquisition of a given area on Earth in a continuous period of time when it passes over this area.
We call this period of time the access period.
However, we restrict ourselves to a discrete set of \textit{imaging attempts} by dividing the access period into multiple equidistant points in time where the satellite can start an acquisition.
The mission plan is restricted to a single orbital revolution so that each acquisition request has at most one access period.
\begin{figure}[t!]
  \centering
  \includegraphics[width=0.65\columnwidth]{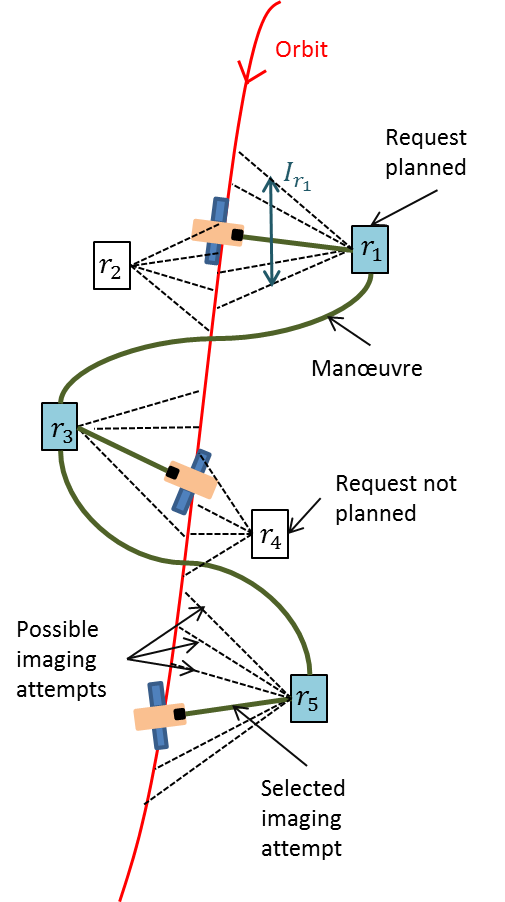}
  \caption{Illustration of a mission plan}
  \label{fig:missionPlanningScheme}
\end{figure}
We strive to select imagining attempts such that the total score is maximized while obeying multiple constraints.
The problem is defined by the following parameters (cf.~Figure~\ref{fig:missionPlanningScheme}):
\begin{itemize}
    \item $R$, the set of acquisition requests in the observed area
    \item $I_r$, the set of imaging attempts for a request $r\in R$ resulting from the discretisation of the access period
    \item $w_{r,i} \in \mathbb{R}$, the score of an imaging attempt $i \in I_r$ for the acquisition request $r \in R$ representing the commercial value of the acquisition
    \item $t^r_i \in \mathbb{R}$, the start time for the acquisition of the imaging attempt $i \in I_r$
    \item $d_{r,i}^\text{acq} \in \mathbb{R}$, the duration of the acquisition of the imaging attempt $i \in I_r$
    \item $d_{(r,i) \to (s,j)}^\text{man} \in \mathbb{R}$, the duration of the manoeuvre from the satellite attitude after finishing acquisition of the imaging attempt $i \in I_r$ to the attitude required to perform the acquisition of the imaging attempt $j \in I_s$.
\end{itemize}
Note, that usually the same score is used for each imaging attempt of a given acquisition request as it corresponds to the commercial value of the acquisition, i.e.\ $w_{r,i} \to w_{r}$.

We define $x_{r,i}$ as the binary variable to capture the selection (value = 1) or dismissal (value = 0) of an acquisition request $r \in R$ at the imaging attempt $i \in I_r$ in the mission plan.
These $x_{r,i}$ are therefore the decision variables of our mission planning problem and their values constitute a mission plan.
Given such a mission plan, the revenue function is the total score for the plan.
Since we would like to formulate the problem as a minimization, we define the cost function as the negative revenue function
\begin{equation}\label{eqn:cost_function}
    C = - \sum_{r \in R} \sum_{i \in I_r} w_{r,i} x_{r,i} \, .
\end{equation}
Not all sets of decision variables constitute a feasible mission plan.
They have to fulfill certain constraints.
First, an acquisition request should be met at most once.
    \begin{equation}\label{eqn:one_attempt_per_request}
        \sum_{i \in I_r} x_{r,i} \leq 1 \quad \forall r \in R.
    \end{equation}
Second, a satellite may not have sufficient time to manoeuvre from the end of a certain image acquisition to the beginning of another one.
Therefore, certain pairs of imaging attempts corresponding to different requests are forbidden to be planned together
    \begin{equation}\label{eqn:forbidden_manoeuvres}
        x_{r,i} + x_{s,j} \leq 1 \quad \forall (i, j) \in F_{r,s} ~\forall (r, s) \in R^2, r\neq s,
    \end{equation}
where
    \begin{align*}
        F_{r,s} = \big\{&(i, j) \in I_r \times I_s : \\ 
                        &t^r_i \leq t^s_j < t^r_i + d_{r,i}^\text{acq} + d_{(r,i) \to (s,j)}^\text{man} \big\} \, 
    \end{align*}
is the set of pairs of forbidden imaging attempts.
The total optimisation problem reads
\begin{equation}
\begin{alignedat}{2}
\label{eqn:classical_modelling}
\text{min } &- \sum_{r \in R} \sum_{i \in I_r} w_{r,i} x_{r,i}\  \\
\text{s.t. } & \sum_{i \in I_r} x_{r,i} \leq 1 &&\forall r \in R \\
            & x_{r,i} + x_{s,j} \leq 1  && \forall (i, j) \in F_{r,s} \\
            &                           && \forall (r, s) \in R^2, r\neq s \\[1ex]
            & x_{r,i} \in \{0, 1\}      &&\forall i \in I_r ~\forall r \in R.
\end{alignedat}
\end{equation}
This problem belongs to the class of integer linear programming (ILP) problems, which are known to be NP-hard in general.
Note, that this problem can be easily extended to a constellation of satellites addressing the same set of imaging requests.

\subsection{Problem instances}
We created sets of benchmark problem instances which are small enough to be amenable to the D-Wave 2000Q machine, but still retain the characteristics of industry size problems.
To this end, we simulated small real-world scenarios and varied a number of scenario parameters to generate a representative set of problem instances.
In particular, we varied
\begin{itemize}
    \item
    the number of acquisition requests in the scenario $N_R = |R|$,
    \item
    the discretisation step $\Delta t$ as the number of seconds between two imaging attempts of the same request
    and 
    \item
    the latitude range $\Lambda$ in degree.  
    The $N_R$ acquisition requests are taken randomly within the latitude interval $[-\Lambda; \Lambda]$ corresponding to the observed part of the orbit track.
\end{itemize}
After creating the instances, it is worthwhile to investigate the derived parameters as they are indicators for the complexity of the instances:
\begin{itemize}
    \item
    number of binary variables $N = \sum_{r\in R} |I_r|$ 
    \item
    constraint ratio (given in percent)
    \begin{equation}
        n_\text{C} = 100 \cdot \frac{2 N_C}{N (N - 1)} \, ,
    \end{equation}
    where 
    \begin{equation}
        N_C = \sum_{\substack{s,r \in R \\ s \neq r}} |F_{r,s}| + \sum_{r \in R} \tfrac{1}{2}|I_r|(|I_r|-1)
    \end{equation}
    is the  number of the pairwise constraints and $\tfrac{1}{2}N {(N - 1)}$ is the maximum number of such constraints.
\end{itemize}

\begin{figure}[h!]
  \centering
  \includegraphics[width=0.49\columnwidth]{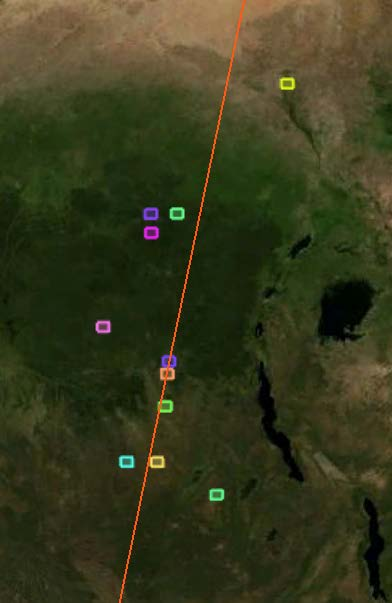} 
  \includegraphics[width=0.49\columnwidth]{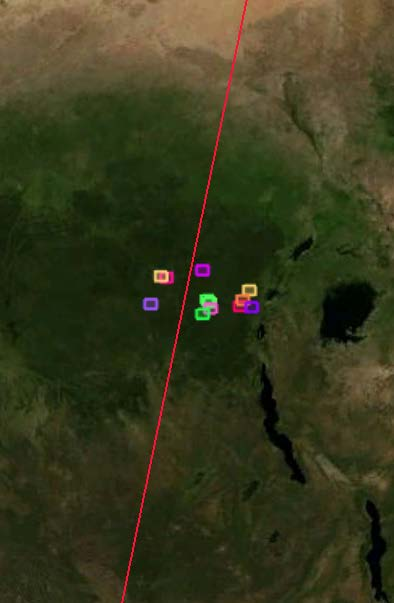}
  \caption{
      Two examples of problem instances used in this study. 
      Left: for $N_R=11$, $\Delta t=12s$, $\Lambda=\ang{10}$ get $N=75$, $n_\text{C}=20\%$.
      Right: for $N_R=12$, $\Delta t=16s$, $\Lambda=\ang{1}$ get $N=68$, $n_\text{C}=42\%$.
  }
  \label{fig:scenarioVisu}
\end{figure}

For simplicity, we set all scores $w_{r,i}=1$.
The number of binary variables $N$
depends on the number of acquisition requests $N_R$ and the number of imaging attempts per request $|I_r|$.
However, the latter is controlled by the discretisation step $\Delta t$.
The finer the discretisation, the more imaging attempts, and therefore binary variables we have.
The constraint ratio $n_\text{C}$ also depends on the latitude range $\Lambda$.
By decreasing the latitude range while fixing all other parameters, the requests get more dense and therefore the time between two requests available for manoeuvres is decreased leading to more constraints.
This can be seen in figure \ref{fig:scenarioVisu} where we show two problem instances with a large and a small latitude range along the same orbit track.

\begin{figure}
\begin{center}
    \includegraphics[width=0.9\columnwidth]{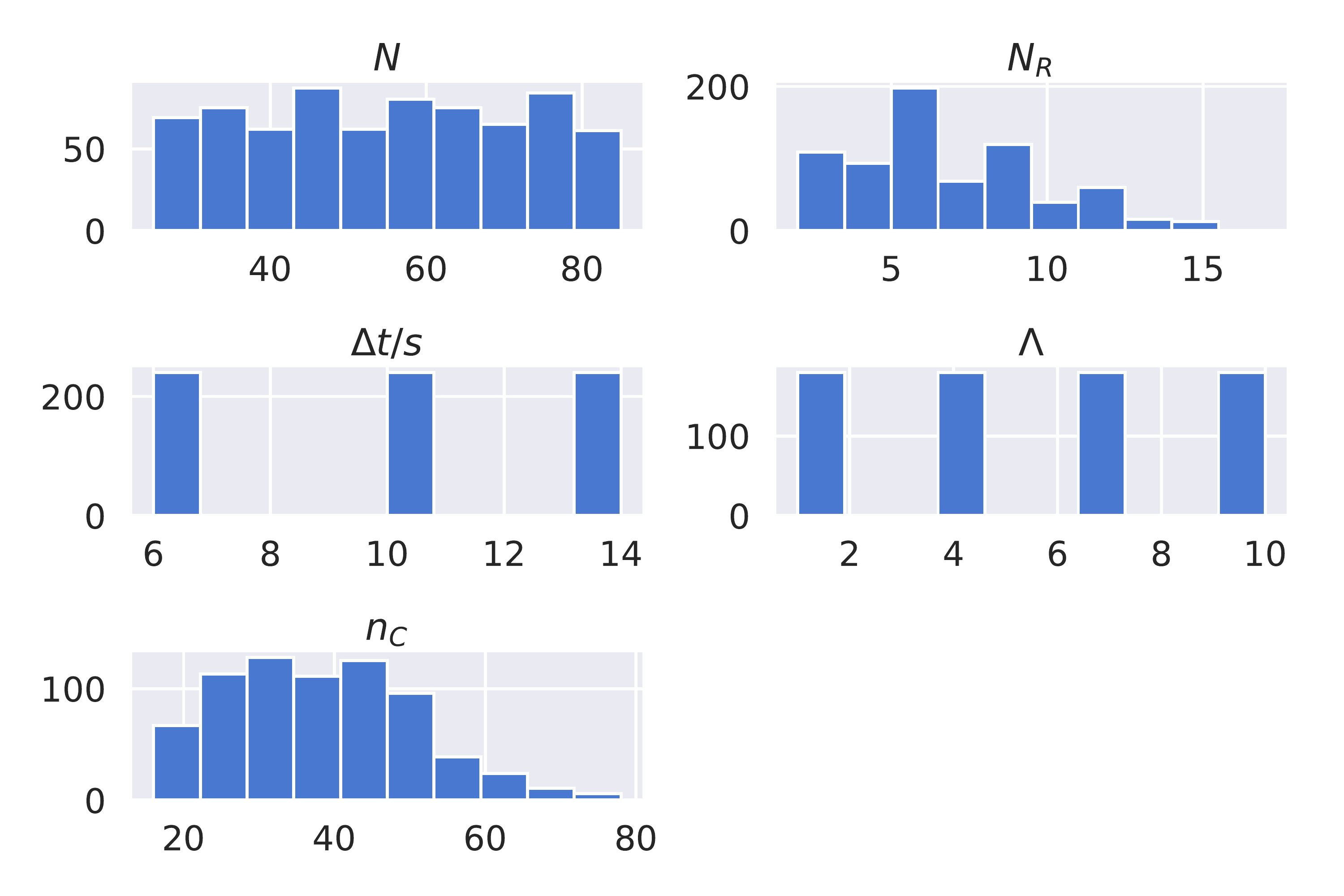}
\end{center}
\caption{Statistics of the set of problem instances $\mathcal{P}$
    showing the amount of instances for different parameters:
    $N$ is the number of binary variables,  $N_R$ is the number of requests, $\Delta t$ is the time discretisation, $\Lambda$ is the latitude range in degrees and $n_C$ is the constraint ratio in percent.
}
\label{fig:instance_statistics}
\end{figure}

We created two sets of problem instances.
The first set $\mathcal{P}$ contains $720$ problem instances with a broad range of parameters. 
Unless stated otherwise, we will use this problem set throughout this work.
Figure \ref{fig:instance_statistics} shows the parameter ranges for the problem set $\mathcal{P}$.
In addition, we created a smaller problem set of $50$ instances with fixed discretisation time $\Delta t = \si{15 s}$ and small latitude range $\Lambda = \ang{2}$, denoted by $\mathcal{P}_\text{hard}$.
This set contains 10 problem instances for each number of binary variables in the list $[30, 40, 50, 60, 70]$.
Due to the small latitude range, these problems are characterised by a relatively large number of constraints which makes them harder to solve in general.

\section{Classical Solution}
\label{sec:classical_solution}
In this section, we discuss two types of classical solution methods: an exact solver and a greedy heuristic approach mimicking the algorithm used operationally at Airbus Defence and Space. 
For both methods, a computer with an Intel(R) Xeon(R) Gold 6152 CPU @ 2.10GHz and 520Go RAM was used.

\subsection{Exact solver}
The open-source, mixed-integer linear programming exact solver \textit{lpsolve} \cite{lpsolve} is used to solve the constraint binary problem \eqref{eqn:classical_modelling}.
Since the formulation of the problem treats constraint \eqref{eqn:forbidden_manoeuvres} in a pairwise fashion, we refer to this method as \textit{pairwise exact solver} hereafter.

In addition, we reformulate problem \eqref{eqn:classical_modelling} to better adapt it for the selected exact solver (lpsolve).
We start by considering a graph consisting of a vertex for each $x_{r,i}$ and an edge between two vertices whenever the activation of both corresponding decision variables is forbidden by one of the two constraints \eqref{eqn:one_attempt_per_request} or \eqref{eqn:forbidden_manoeuvres}.
As a pre-processing step, we find the set of all maximal cliques $S_\text{maxcliques}$ in this graph. 
The mission planning problem \eqref{eqn:classical_modelling} can then be reformulated as:
\begin{equation}
\begin{alignedat}{2}
\label{eqn:classical_modelling_cliques}
\text{min }   & - \sum_{r \in R} \sum_{i \in I_r} w_{r,i} x_{r,i}\  \\
\text{s.t } &\sum_{x_{r, i} \in K} x_{r,i} \leq 1 &&\forall K \in S_\text{maxcliques}\\
            & x_{r,i} \in \{0, 1\}      &&\forall i \in I_r ~\forall r \in R.\\
\end{alignedat}
\end{equation}
The equivalence of this problem to \eqref{eqn:classical_modelling} can be demonstrated by contradiction. 
Since maximal cliques can have better properties for exact solving \cite{Gabrel2010, Gabrel2006, Kiatmanaroj2013}, we expect the exact solver to perform better on the reformulated problem than on the original one.
For the remainder of this work, we refer to the method of solving the reformulated problem \eqref{eqn:classical_modelling_cliques} with lpsolve as \textit{clique exact solver}.

\subsection{Greedy heuristic}
\label{sec:greedy}
Due to run-time requirements, a greedy algorithm with complexity $\mathcal{O}(N)$ is used operationally. 
The algorithm considers all acquisition requests one by one in decreasing order of score. The algorithm tries to insert each acquisition request in the plan (initially empty) while preventing manoeuvring conflicts. Higher value requests therefore benefit from a higher likelihood of being inserted into the plan, due to the relative sparsity of entries in the early parts of the planning. Lower value requests, considered only at the end of the planning process, are more difficult to insert into the plan while respecting all manoeuvring constraints.
In order to find at least one imaging attempt for the considered request which does not violate any manoeuvring constraint involving the already inserted requests, the algorithm is permitted to change their corresponding selected imaging attempts.
If no suitable imaging attempt can be found, the request is discarded.
The decision of inserting or discarding an acquisition request is final and will not change when trying to insert following lower-score requests.
The chronological order of the acquisition requests inserted in the plan never changes throughout the algorithm.
The algorithm stops when all requests have been considered.

In most problems this deterministic algorithm will not provide the optimal solution but it is a good compromise between a fast execution time and high total score value.

\section{Quantum Annealing}
\label{sec:quantum_annealing}
In order to make the problem amenable to a D-Wave quantum annealer, we need to convert its mathematical formulation to a quadratic unconstrained binary optimisation (QUBO) formulation \cite{rieffel2015,Venturelli2015,stollenwerkATM2019}.
This can be done by adding penalty terms to the linear cost function \eqref{eqn:cost_function} in order to enforce the constraints.
The total QUBO cost function is then given by
\begin{equation} \label{eqn:qubo}
    Q = C + \lambda_u C_u + \lambda_t C_t \, .
\end{equation}
The first constraint \eqref{eqn:one_attempt_per_request} is fulfilled if the penalty term
\begin{equation}\label{eqn:qubo_one_attempt_per_requests}
	C_u = \sum_{r \in R} \sum_{\substack{i, j \in I_r \\ i \neq j}} x_{r,i}x_{r,j} 
\end{equation}
vanishes.
Similarly, the second constraint \eqref{eqn:forbidden_manoeuvres} is fulfilled if the penalty term
\begin{equation}\label{eqn:qubo_forbidden_manoeuvres}
    C_t = \sum_{\substack{r,s \in R \\ r \neq s}} \sum_{(i, j) \in F_{r,s}} x_{r,i} x_{s,j} 
\end{equation}
vanishes.
In order to make sure that both penalty terms vanish in the optimal solution of the problem, we need to choose the penalty weights $\lambda_t$ and $\lambda_u$ large enough. 
An upper bound for the minimum sufficient penalty weights is given by
\begin{equation}
    \overline{\lambda} = \max_{\substack{r \in R \\ i \in I_r}} w_{r,i} \, .
\end{equation}
Meaning, if $\lambda_t > \overline{\lambda}$ and $\lambda_u > \overline{\lambda}$, the optimal solution fulfils the constraints.
Therefore, we chose 
\begin{equation}\label{eqn:penalty_weights}
    \lambda_u = \lambda_t = 1.1 \overline{\lambda}
\end{equation}
for the remainder of this work.

\begin{figure}
\begin{center}
    \includegraphics[width=1.0\columnwidth]{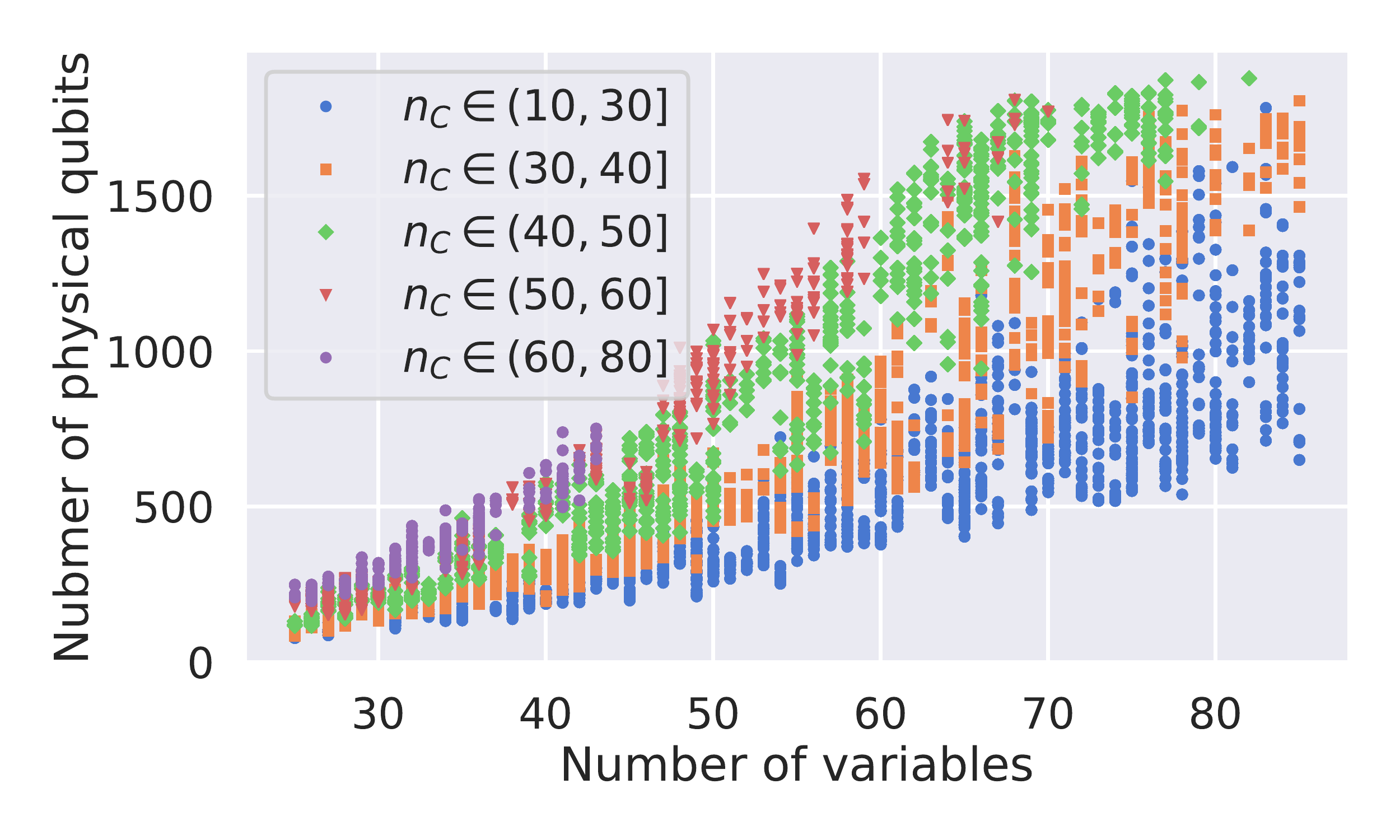}
\end{center}
\caption{Number of physical qubits after embedding against number of variables (number of logical qubits) for various connectivity ratios $n_c$ for all instances in $\mathcal{P}$.}
\label{fig:embedding}
\end{figure}

The D-Wave 2000Q processor has a \textit{Chimera}-graph architecture. In order to map problems with arbitrary connectivity onto the machine, it is necessary to first convert the problems into a \textit{Chimera}-graph form \cite{Choi08,cai2014}.
For each instance in our problem set, we use up to 5 different solutions of D-Wave's heuristic embedding algorithm \cite{cai2014} to embed a given instance into the D-Wave 2000Q processor.
We were able to embed problem instances with up to 80 logical variables (see figure \ref{fig:embedding}).
As expected, problems with a larger number of connections between imaging requests need a larger number of physical qubits per logical qubits.

We used $10000$ annealing runs and majority voting as an unembedding strategy for each of the embedded problem instances.
For each instance the chain coupling $J_C$ was fixed to a constant value chosen in two different ways (cf.~\cite{stollenwerkATM2019,stollenwerkFGA2019}): 
\begin{enumerate}[(a), parsep=0pt]
    \item by taking the negative magnitude of the largest absolute coefficient value of the corresponding Ising model (converted from QUBO) before the embedding. We will call this \textit{worst case treatment}. 
    \item by experimentally finding the value that provides the largest success probability, which we will call \textit{optimal-value}. 
\end{enumerate}
The success probability is measured by dividing the number of optimal solutions found by the number of annealing runs \cite{rieffel2015,Venturelli2015}.
We remark that the \textit{optimal-value} strategy is poorly suited for use in an operational setting due to the repeated calls to the D-Wave machine during parameter optimisation.

\begin{figure}
\begin{center}
    \includegraphics[width=1.0\columnwidth]{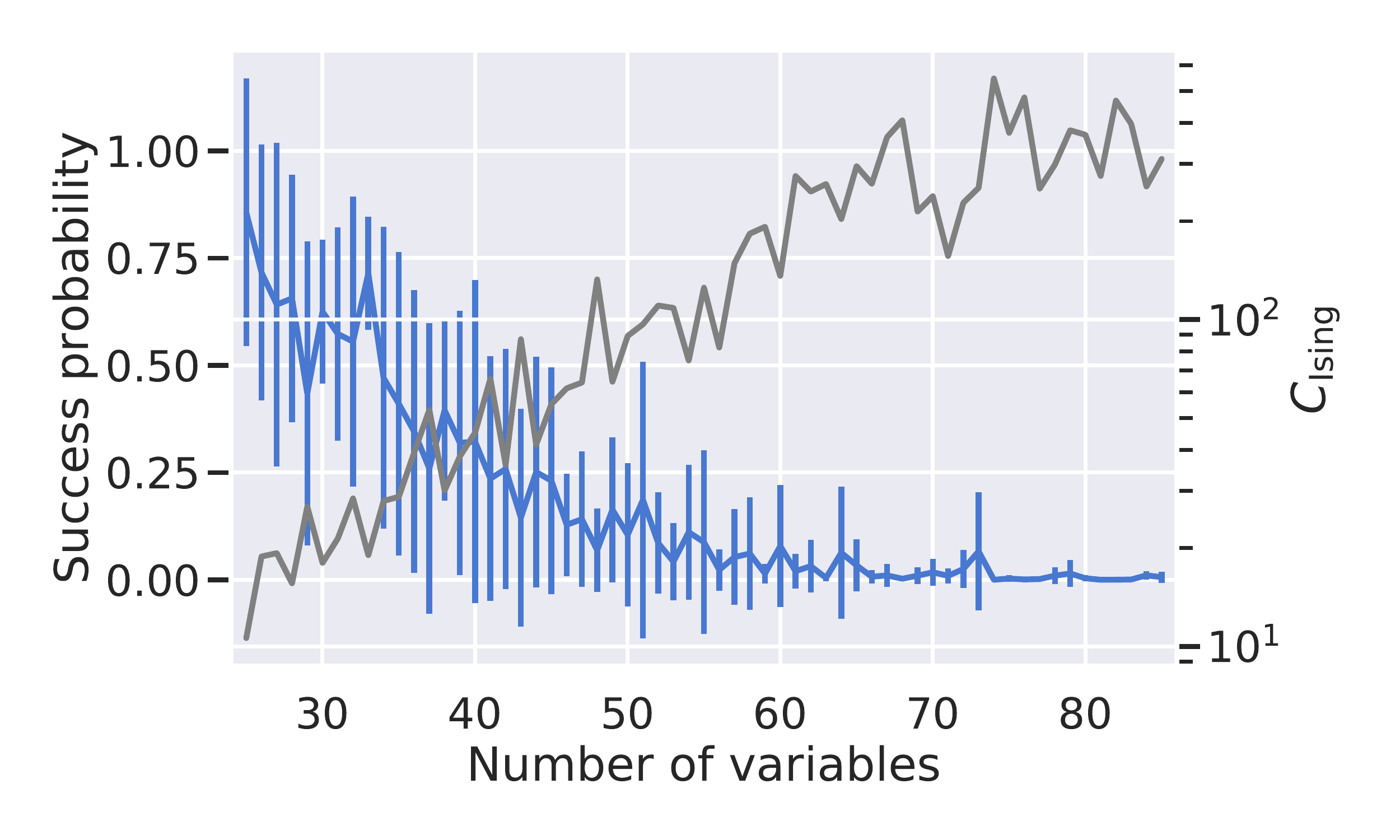}
\end{center}
\caption{Success probability (mean and standard deviation, blue) and the maximum coefficient ratio (mean, grey) against the number of variables (number of logical qubits) for all instances in $\mathcal{P}$. The chain coupling was set experimentally for each instance in order to maximise the success probability.}
\label{fig:successProbability}
\end{figure}

As was shown in previous studies, the precision requirement of the problem instances can be a limiting factor \cite{stollenwerkATM2019,stollenwerkFGA2019}.
Given the embedded Ising model of a certain problem instance 
$H = \sum_i h_i s_i + \sum_{ij} J_{ij} s_i s_j, \; s_i \in \{-1, 1\}$,
a measure of the required precision is the maximum coefficient ratio (cf.~\cite{stollenwerkATM2019,stollenwerkFGA2019}):
\begin{equation}
    C_\text{Ising} = \max\left\{\frac{\max_i | h_i |}{\min_i | h_i |} ,\frac{\max_{ij} | J_{ij} |}{\min_{ij} | J_{ij} |} \right\} \, .
\end{equation}
If this number is large, the problem cannot be resolved on the D-Wave machine and the success probability is suppressed. 
Figure \ref{fig:successProbability} shows the success probability and the maximum coefficient ratio for all instances against the number of variables.
Although, we were able to embed instances up to 80 logical variables, the success probability vanishes for instances larger than approximately 65.
Simultaneously, the maximum coefficient ratio, and therefore the precision requirements, increases exponentially with the number of logical variables up to a plateau above 65.
From these results, we conclude that the precision is the limiting factor here.

\section{Quantum Annealing vs. Classical Resolution}
\label{sec:qa_vs_classical_comparison}
\subsection{Random sampling vs QA}
\label{sec:random_vs_qa}
\begin{figure}[t]
  \centering
  \includegraphics[width=1.0\columnwidth]{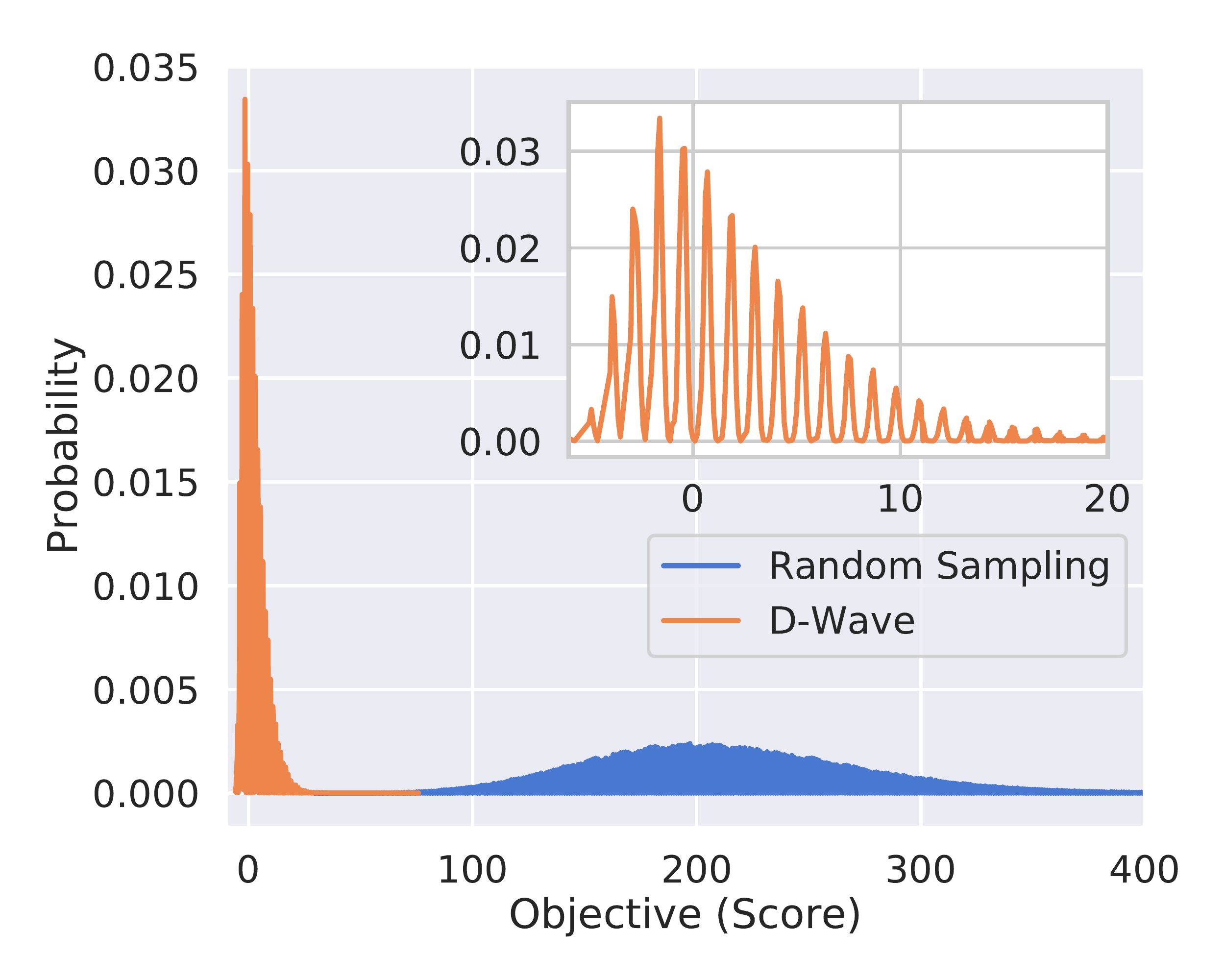}
  \caption{Random sampler against the D-Wave machine for QUBO resolution of an example problem instance with the following characteristics: $N=70$, $\Delta t=15s$, $\Lambda=\ang{2}$, ${N_R=12}$, $n_C=37\%$.}
  \label{fig:pdf_dwave}
\end{figure}

As a first step in our comparison of classical methods and quantum annealing, we consider the QUBO cost function of a single problem instance and compare the energy distribution obtained with the quantum annealer against one obtained with random uniform sampling of the search space for our binary variables.
The result can be seen in figure \ref{fig:pdf_dwave}.
We used 100000 quantum annealing runs on the chosen problem instance with 70 binary variables.
As expected, the quantum annealer samples from the low energy distribution of the given QUBO and gives far better results than a random uniform sampling.

The oscillatory behaviour in the D-Wave or the random evaluation is caused by the encoding approach of constraints. 
The oscillation period is equal to the value of the selected penalty weights: $\lambda_t = \lambda_u = 1.1\overline{\lambda}$ (cf.\ equation \eqref{eqn:penalty_weights}).

\subsection{Time-to-exact-solution benchmark}
\begin{figure}[t]
  \centering
  \includegraphics[width=1.0\columnwidth]{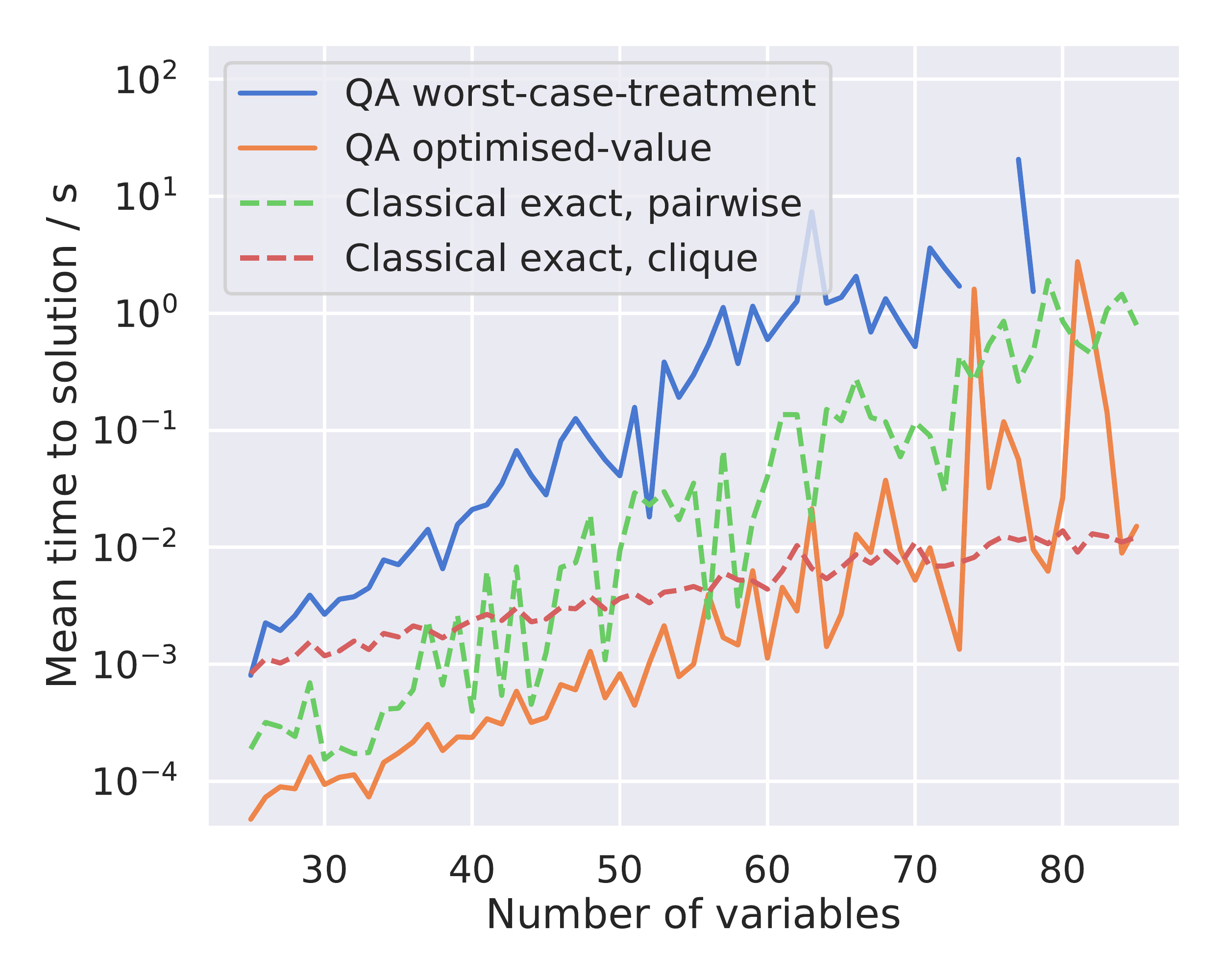}
  \caption{Run time comparison of classical exact solver and quantum annealer. 
      For quantum annealing, the time to exact solution with 99\% certainty is given.
  }
  \label{fig:runtime_comparison}
\end{figure}

In this section, we compare the time needed for the exact solver to find the exact solution with the time needed for the QA to find the optimal solution with $99\%$ certainty. 
The latter is given in terms of the success probability $p$ as
\begin{equation*}
    T_{99} = \frac{\ln(1 - 0.99)}{\ln(1 - p)} T_\text{Anneal} \, ,
\end{equation*}
where the annealing time was set to $T_\text{Anneal}=20\mu s$ for all experiments.
As discussed in section \ref{sec:quantum_annealing}, we used the two different choices of the chain coupling $J_C$ denoted by \textit{worst-case-treatment} and \textit{optimised-value}.
Moreover, we used two different classical exact methods, denoted by \textit{pairwise exact solver} and \textit{clique exact solver} (cf.~section~\ref{sec:classical_solution}).
The result of this comparison is shown in figure \ref{fig:runtime_comparison}, where the run time is averaged over all problem instances having the same number of binary variables. 
As expected, the classical execution time of the \textit{pairwise exact solver} increases exponentially with the number of binary variables. 
It is not surprising as ILPs in general belong to the complexity class of NP-hard problems. 
The quantum annealing results with \textit{worst-case-treatment} show a similar slope and a constant offset of about one order of magnitude.
By optimising the coupling chain strength (\textit{optimised-value}), quantum annealing runs faster than the \textit{pairwise exact solver}.
However, \textit{clique exact solver} performs better than all other methods for larger instances.
From this we cannot conclude however, that the quantum annealer is inferior to classical methods in general.
As explained in section \ref{sec:quantum_annealing}, the precision problems suppress the success probability significantly for moderately large problem instances.

\subsection{Quality of solution benchmark}
In an operational setting, it is often the case that a fixed time budget is reserved for the solver.
The goal is then to find the best solution in a fixed time frame.
In order to investigate the performance of a quantum annealer in such a setting, we fixed the execution time for the quantum annealer and compared it to the greedy heuristic described in section \ref{sec:greedy}.
A fixed execution time for the quantum annealer means we restrict the number of annealing runs $n$ to a certain value.
Then the execution time is given by $n T_\text{anneal}$.
Where we chose $T_\text{anneal}=20\mu s$.
Note that we neglect pre- and post-processing time for the quantum annealer.
As a measure for the solution quality, we use the approximation ratio,
i.e.\ the objective value of the best found solution divided by the optimal objective value.

\begin{figure}
  \centering
  \includegraphics[width=1.0\columnwidth]{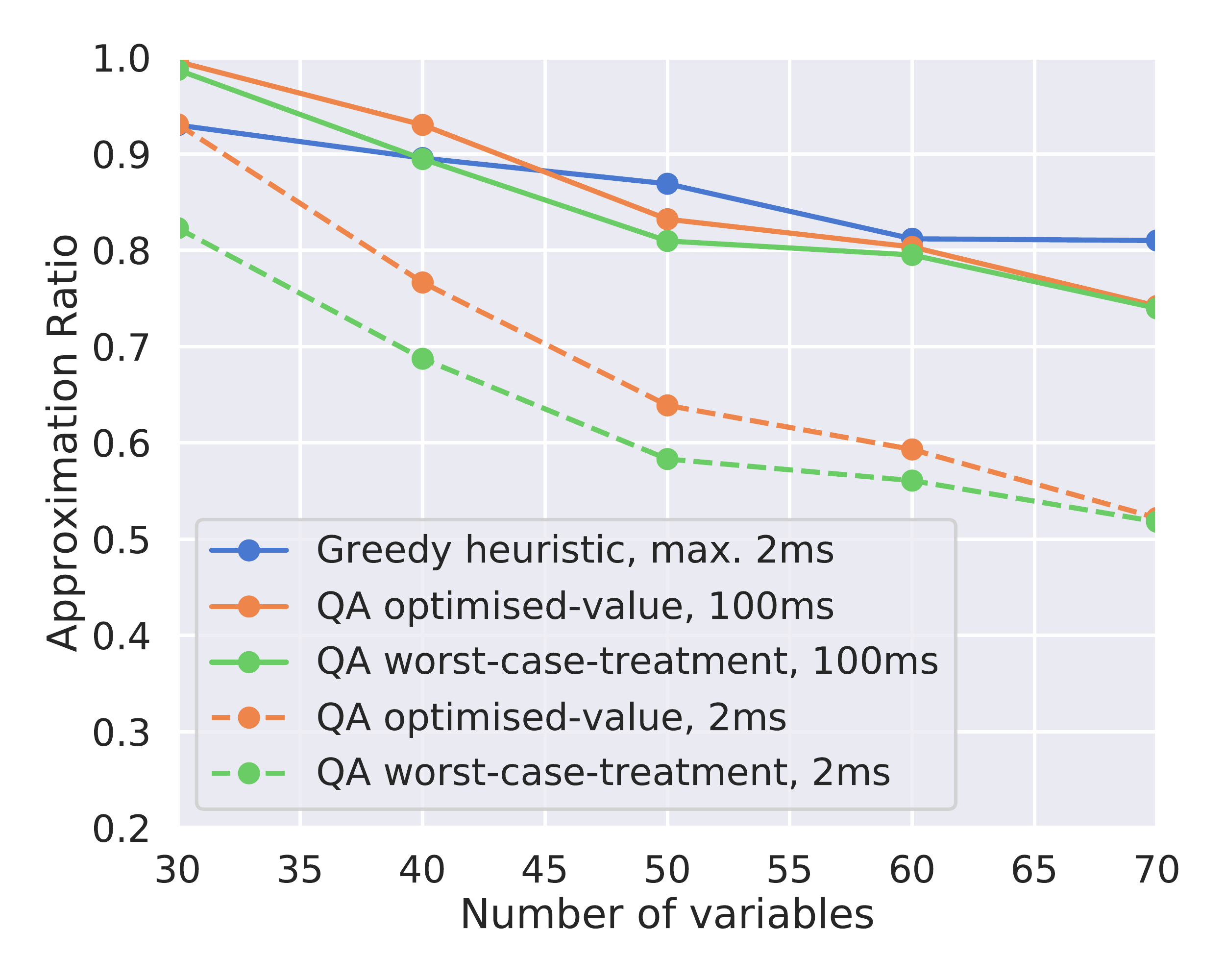}
  \caption{Classical greedy heuristic vs QA for quality of solution with fixed time.
        Orange: optimised chain coupling, Green: fixed chain coupling, cf.\ \ref{sec:quantum_annealing}.
    }
  \label{fig:benchmarkFixedTime}
\end{figure}
Figure \ref{fig:benchmarkFixedTime} shows the average approximation ratio against the number of variables for the greedy heuristic as well as various quantum annealing configurations,
for the problem set $\mathcal{P}_\text{hard}$. 
For the annealing runs, we considered both choices of the chain strength and different fixed execution times.
The Greedy algorithm execution was fixed to maximally \si{2 ms}.
It is obvious that the greedy heuristic outperforms the quantum annealer for similar execution times.
Only for larger execution times, the quantum annealer yields better results than the greedy heuristic for smaller instances.
However, due to the precision issues of the quantum annealer the performance is suppressed for larger instances.

\section{Conclusion}
\label{sec:conclusion}
In this paper, we have studied a real-world satellite planning problem and benchmarked the D-Wave 2000Q quantum annealer against classical solvers, using exact algorithms as well as the heuristic algorithm implemented in operational systems. We have detailed the steps required to derive a suitable formulation for the D-Wave machine and investigated different approaches for setting the chain coupling factor.

As expected, the D-Wave quantum annealer samples the low energy states of our problem far more efficiently than a random sampler. We have proposed a fair methodology to evaluate quantum annealing against classical computing through two benchmarks: time to exact solution and quality of solution with fixed time. Results show similar trends for both technologies but do not reveal any advantage over classical computers for this particular problem.

Limited qubit connectivity, precision issues and coherence time remain a major bottleneck for the D-Wave 2000Q processor. Nevertheless, quantum annealers can already compete with classical solutions that benefit from decades of research and continuous improvement on both the computing hardware and the optimization software. 
We expect future quantum annealing devices to improve significantly in size and accuracy.
Based on the results of this study, we believe such future devices could outperform classical computers for industry size problems.
Additional studies similar to this one are likely to enable the use of more capable quantum computers for real-world applications.



\section{References}
\bibliographystyle{abbrv}
\bibliography{references}

\end{document}